# From the AI Act to a European AI Agency: Completing the Union's Regulatory Architecture

Georgios Pavlidis

More information about the journal and submission process can be found at
https://www.cyelp.com/index.php/cyelp/about

# FROM THE AI ACT TO A EUROPEAN AI AGENCY: COMPLETING THE UNION'S REGULATORY ARCHITECTURE

Georgios Pavlidis[*]

*Abstract:* As artificial intelligence (AI) technologies continue to advance, effective risk assessment, regulation, and oversight are necessary to ensure that AI development and deployment align with ethical principles while preserving innovation and economic competitiveness. The adoption of the EU AI Act marks an important step in this direction, establishing a harmonised legal framework that includes detailed provisions on AI governance, as well as the creation of the European AI Office. This paper revisits the question of whether a more robust supranational agency dedicated to AI is still warranted and explores how such a body could enhance policy coherence, improve risk assessment capacities, and foster international cooperation. It also argues that a strengthened EU-level agency would also serve the Union's strategic aim of securing digital and technological sovereignty.

## 1 Introduction

Artificial intelligence (AI) technologies and applications are rapidly expanding and being commercialised across numerous sectors, including healthcare, commerce, transportation, finance, and beyond.[1] Nevertheless, the growth of AI has also given rise to serious ethical, societal, and legal challenges (section 2).[2] For this reason, effective risk assessment, regulation, and oversight of AI are required to ensure that this growth aligns with ethical principles without hindering innovation (section 3). The objective of this paper is to examine whether the establishment of a supranational agency on AI would fit this imperative and how the current European Union (EU) legal and institutional framework—particularly following the establishment of the EU AI Office— needs to be reinforced to address AI risks (section 4).

---

[*] Associate Professor of International and EU Law, Director, Jean Monnet Centre of Excellence 'AI-2-TRACE-CRIME' Neapolis University Pafos (NUP), Cyprus. This research was funded by the European Union. However, the views and opinions expressed are those of the author only and do not necessarily reflect those of the European Union or the European Education and Culture Executive Agency (EACEA). Neither the European Union nor EACEA can be held responsible for them. ORCID: 0000-0001-6311-3086; DOI: 10.3935/cyelp.21.2025.610.

[1] Commission, 'Building Trust in Human Centric Artificial Intelligence' (Communication) COM (2019) 168 final.

[2] Organization for Economic Co-operation and Development, 'Advancing Accountability in AI: Governing and Managing Risks Throughout the Lifecycle for Trustworthy AI' (2023) 349 OECD Digital Economy Papers 26; Michael Littman and others, 'Gathering Strength, Gathering Storms: The One Hundred Year Study on Artificial Intelligence (AI100), 2021 Study Panel Report' (Stanford University, September 2021).

Many jurisdictions have taken increasing steps to regulate AI,[3] but disparities exist in their approaches to AI standards and enforcement mechanisms. This fragmentation is suboptimal because AI products and services transcend national borders, and attempting to regulate a global issue at multiple local levels has consistently proven to be an uphill battle.[4] Even where effective norms are adopted at the national level, enforcing them in a borderless and interconnected digital environment will be a challenge.[5] International organisations such as the Organization for Economic Co-operation and Development (OECD) and the United Nations Educational, Scientific and Cultural Organization (UNESCO) have proposed guidelines,[6] but they lack binding regulatory power. For its part, the recently adopted Council of Europe's Framework Convention on AI faces significant challenges on the path to effective implementation and, by its nature, cannot achieve global applicability.[7] Thus, disparities between national legal frameworks on AI are very likely to remain unaddressed. The lack of a global framework is likely to complicate compliance and hinder innovation, as legal uncertainty and varying regulatory requirements will persist across jurisdictions.[8] Although fragmentation may sometimes be creative in global governance,[9] it must be avoided in the context of AI regulation, as it might impede the effective management of the significant societal impacts of AI, including issues related to bias, accountability, and transparency.

---

[3] In addition to the EU action, which is discussed in this paper, see the legislative initiatives on AI in Brazil (Projeto de Lei n° 2338, de 2023), in China (2021 Regulation on Recommendation Algorithms; 2022 Rules for Deep Synthesis; 2023 Draft Rules on Generative AI), and in Canada (Draft Law C-27, Digital Charter Implementation Act 2022, Part 3: Artificial Intelligence and Data Act). The United Kingdom intends to enhance the responsibilities of its current regulatory bodies, such as the Information Commissioner's Office, the Financial Conduct Authority, and the Competition and Markets Authority, rather than implementing comprehensive new legislation to regulate AI, which differs from the approach taken by the EU. These organisations will have authority to offer guidance and supervise the utilisation of AI within their specific domains of jurisdiction; UK Secretary of State for Science, Innovation and Technology, 'A Pro-Innovation Approach to AI Regulation' (Policy Paper, 2023) <www.gov.uk/government/publications/ai-regulation-a-pro-innovation-approach/white-paper> accessed 14 September 2025.

[4] Jonathan Wiener, 'Think Globally, Act Globally: The Limits of Local Climate Policies' (2007) 155 University of Pennsylvania Law Review 1961; James Bushnell, Carla Peterman, and Catherine Wolfram, 'Local Solutions to Global Problems: Climate Change Policies and Regulatory Jurisdiction' (2008) 2(2) Review of Environmental Economics and Policy 175.

[5] Julia Hörnle, *Internet Jurisdiction - Law and Practice* (OUP 2021). This holds especially true within the realm of cloud computing. On this issue, see Christopher Millard (ed), *Cloud Computing Law* (2nd edn, OUP 2021) and Dinesh Soni and Neetesh Kumar, 'Machine Learning Techniques in Emerging Cloud Computing Integrated Paradigms: A Survey and Taxonomy' (2022) 205 Journal of Network and Computer Applications 103419. See also Isabelle Bousquette, 'The AI Boom Is Here - The Cloud May Not Be Ready' *The Wall Street Journal* (10 July 2023).

[6] OECD, Recommendation of the Council on Artificial Intelligence, OECD/LEGAL/0449, 22 May 2019; UNESCO, Recommendation on the Ethics of Artificial Intelligence, Document No SHS/BIO/PI/2021/1, 2021.

[7] The Council of Europe's Committee of Ministers had mandated the Committee on Artificial Intelligence (CAI) to elaborate a framework Convention on the development and application of AI, based on the standards of the Council of Europe <www.coe.int/en/web/artificial-intelligence> accessed 20 October 2025.

[8] Peter Cihon, Matthijs Maas, and Luke Kemp, 'Fragmentation and the Future: Investigating Architectures for International AI Governance' (2020) 11(5) Global Policy 545.

[9] Amitav Acharya, 'The Future of Global Governance: Fragmentation May be Inevitable and Creative' (2016) 22(4) Global Governance 453.

For the EU, the adoption of the AI Act[10] constitutes an important milestone in establishing a harmonised and risk-based regulatory framework for AI across the 27 Member States. However, despite this important step, challenges remain regarding the standard-setting, as well as the supervision and enforcement of the rules on AI. The EU AI Office—established under the AI Act—is intended to perform many coordination and oversight functions. However, its powers and structure are more limited than what might be envisioned for a full-fledged supranational agency. This limitation is particularly significant given the AI Act's dual character: while framed as economic regulation to secure the internal market, its provisions fundamentally address fundamental rights, safety, and broader societal risks.[11] This expansive scope distinguishes it from traditional product-safety legislation and strengthens the case for a more specialised supranational body with the stronger oversight powers necessary to protect citizens, not just markets. A logical next step in the EU's regulatory trajectory would be the consolidation of AI governance functions within a dedicated and strengthened EU agency on AI. Such an agency could serve as a central authority to support standard-setting, enforcement of the AI Act, and cross-border coordination. If properly designed, it would strengthen Europe's ability to promote innovation and uphold its strategic goal of technological sovereignty, while also safeguarding fundamental rights. As this paper will argue, advancing this institutional development is timely and necessary given the fast-paced evolution of AI.

## 2 The thrills and perils of AI's ascendance

There are estimates that the global AI market, valued at $428 billion in 2022, will grow to more than $2 trillion by 2030.[12] By then, it is also anticipated that AI will contribute $15.7 trillion to the worldwide economy or a 14% increase in global gross domestic product (GDP), surpassing the present collective economic output of China and India.[13] AI-driven technologies also promise to enhance labour productivity by as much as 40% across a spectrum of 16 industries by 2035.[14] Not surprisingly, the business world, especially Big Tech, is experiencing AI fever and the majority of AI investment now comes from private sources. In 2023, the United States led with €62.5 billion in private AI investment, followed by China with €7.3 billion; that same year, the EU and the United Kingdom combined attracted €9 billion in private

---

[10] Regulation (EU) 2024/1689 of the European Parliament and of the Council of 13 June 2024 laying down harmonised rules on artificial intelligence and amending Regulations (EC) No 300/2008, (EU) No 167/2013, (EU) No 168/2013, (EU) 2018/858, (EU) 2018/1139 and (EU) 2019/2144 and Directives 2014/90/EU, (EU) 2016/797 and (EU) 2020/1828 (Artificial Intelligence Act) [2024] OJ L 2024/1689. See also European Commission, Proposal for a Regulation laying down harmonized rules on artificial intelligence (Artificial Intelligence Act), COM(2021) 206 final.
[11] Marco Almada and Nicolas Petit, 'The EU AI Act: Between the Rock of Product Safety and the Hard Place of Fundamental Rights' (2025) 62 Common Market Law Review 85.
[12] Fortune Business Insights, 'Artificial Intelligence Market' (Market Research Report, April 2023).
[13] PWC, 'Sizing the Price: What's the Real Value of AI for Your Business and How Can You Capitalise?' (PWC Report 2017).
[14] Accenture, 'Why Artificial Intelligence is the Future of Growth' (Accenture Report 2016).

investment in the sector.[15] At the business level, the number of companies using AI capabilities (natural-language generation, robot process automation, computer vision, etc) had increased to 78% by October 2025, up from 55% percent in 2023,[16] while an astonishing 83% of companies view the incorporation of AI into their business strategies as a priority.[17]

The capabilities of AI justify this excitement within the business world, but concerns arise about the associated risks. Historically, this has been true for most emerging novel technologies,[18] but in the context of AI, many of these concerns appear to be well founded. Indeed, ethical dilemmas become more pronounced as AI systems increasingly influence decisions with consequences for humans, especially in sensitive domains such as healthcare, credit scoring, policing, and the criminal justice system.[19] In these domains, the potential for bias and discrimination within AI systems could inadvertently perpetuate preexisting inequalities.[20] This raises issues such as fairness, responsibility, openness, and how these principles should be incorporated into legal frameworks. Furthermore, security concerns take centre stage as the growing dependence on AI creates fresh opportunities for cyberattacks, data compromises, and other malicious uses and abuses of AI.[21] In addition to these difficulties, the consolidation of power and data within a small group of dominant tech companies or entities gives rise to concerns regarding the possible abuse of this authority in markets, public discussions, and even political procedures.[22] Finally, the looming prospect of job displacement is significant, with automation and AI-driven processes posing a threat in terms of reshaping industries, amplifying disparities in the labour market, and profoundly transforming the employment landscape.[23]

---

[15] European Parliament, 'AI investment: EU and global indicators' (European Parliamentary Research Service, March 2024)

[16] McKinsey, 'The State of AI: How Organizations Are Rewiring to Capture Value' (McKinsey Survey, March 2025).

[17] Falon Fatemi, '3 Ways Artificial Intelligence Is Transforming Business Operations' *Forbes* (29 May 2019).

[18] Bernard Cohen, 'The Fear and Distrust of Science in Historical Perspective' (1981) 6(3) Science, Technology, & Human Values 20; Marita Sturken, Douglas Thomas, and Sandra Ball-Rokeach, *Technological Visions: Hopes and Fears That Shape New Technologies* (Temple University Press 2004).

[19] Jacob O Arowosegbe, 'Data Bias, Intelligent Systems and Criminal Justice Outcomes' (2023) 31(1) International Journal of Law and Information Technology 22; Abdul Malek, 'Criminal Courts' Artificial Intelligence: The Way It Reinforces Bias and Discrimination' (2022) 2 AI and Ethics 233; Michael Bücker and others, 'Transparency, Auditability, and Explainability of Machine Learning Models in Credit Scoring' (2022) 73 Journal of the Operational Research Society 70; Georgios Pavlidis, 'Deploying Artificial Intelligence for Anti-money Laundering and Asset Recovery: The Dawn of a New Era' (2023) 26(7) Journal of Money Laundering Control 155.

[20] This may be due to several factors, such as biased training data, data collection methods, feature selection, and feedback loops. See Frederik Zuiderveen Borgesius, 'Discrimination, Artificial Intelligence, and Algorithmic Decision-Making' (Council of Europe Study 2018).

[21] European Union Agency for Cybersecurity, 'Artificial Intelligence Cybersecurity Challenges' ENISA Report (2020); Europol, 'Malicious Uses and Abuses of Artificial Intelligence' Europol Report (2021); on computer-related crimes and virtual criminality, see Ian Lloyd, *Information Technology Law* (9th edn, OUP 2020).

[22] Nick Srnicek, 'Platform Monopolies and the Political Economy of AI' in John McDonnell (ed), *Economics for the Many* (Verso 2018) 152–163; Pieter Verdegem, 'Dismantling AI Capitalism: The Commons as an Alternative to the Power Concentration of Big Tech' (2022) AI & Society <https://doi.org/10.1007/s00146-022-01437-8> accessed 14 September 2025.

Nevertheless, identifying the risks is usually much easier than building consensus on the necessary policy and legal responses.[24] For example, although there is agreement on the need to address the ethical aspects of AI decision-making, policy responses require a prior and clear definition of fairness and bias.[25] Such a definition might be highly dependent on context, leading to differing interpretations, while legal definitions of bias may also differ from one jurisdiction to another. Therefore, any attempt to regulate AI must overcome serious obstacles, such as the definition of AI systems, the criteria for the classification of risks, as well as the scope and criteria of transparency, explainability, and due diligence obligations that will be imposed on AI developers and users.[26]

## 3 The shift towards enforceable AI regulations and supervision

Several public and private organisations have developed or are developing sets of soft-law principles for AI governance,[27] attempting to balance technological innovation with responsible AI. A notable example is the set of non-binding guidelines adopted by the EU in 2019, which are closely related to the principles adopted by the OECD some weeks later in the same year.[28] Nevertheless, the soft-law approach to AI appears to be gradually giving rise to a more robust legislative approach, not only in the EU with its ambitious AI Act but also in other jurisdictions.[29] In this 'race to AI regulation',[30] jurisdictions are seeking to develop new binding rules on AI, supported by enforcement mechanisms. The choice between hard and soft law has attracted scholarly attention across various domains of law and governance.[31] We contend that the transition from soft to hard law is both timely and suitable in the context of AI, given the magnitude of the risks associated with this emerging technology. This does not mean that soft-law rules should be discarded as a policy tool; they provide the benefits of flexibility and adaptability and can be utilised to complement hard law, serving as mutually reinforcing components in AI regulation.[32] Nevertheless, there is a need for a set of binding regulations, with enforcement mechanisms and sanctions as a deterrent against non-compliance.[33]

---

[23] OECD, 'Artificial Intelligence and Employment' (Policy Brief 2021); see also Accenture, 'A New Era of Generative AI for Everyone' (Accenture Report 2023). According to this report, 40% of all working hours can be impacted by Large Language Models (LLMs) like GPT-4.
[24] Charlotte Stix and Matthijs Maas, 'Bridging the Gap: The Case for an Incompletely Theorized Agreement on AI Policy' (2021) 1 AI and Ethics 261.
[25] Vincent Müller, 'Ethics of Artificial Intelligence and Robotics' in Edward Zalta (ed), *Stanford Encyclopedia of Philosophy* (Stanford University 2020).
[26] Leilani Gilpin and others, 'Explaining Explanations: An Overview of Interpretability, of Machine Learning' (2019) ArXiv <http://arxiv.org/abs/1806.00069> accessed 14 September 2025.
[27] Ryan Budish, 'AI's Risky Business: Embracing Ambiguity in Managing the Risks of AI' (2021) 16 Journal of Business & Technology Law 259.
[28] Commission, Ethics Guidelines for Trustworthy AI, High-Level Expert Group on AI, 8 April 2019; OECD, Recommendation of the Council on Artificial Intelligence, OECD/LEGAL/0449, 22 May 2019.
[29] Anu Bradford, 'The Race to Regulate Artificial Intelligence: Why Europe Has an Edge Over America and China' *Foreign Affairs* (27 June 2023); see also the legislative initiatives described in n 3.
[30] Nathalie Smuha, 'From a Race to AI to a Race to AI regulation: Regulatory Competition for Artificial Intelligence' (2021) 13(1) Law, Innovation and Technology 57.
[31] Christine Chinkin, 'The Challenge of Soft Law: Development and Change in International Law' (1989) 38(4) International & Comparative Law Quarterly 850; Kenneth Abbott and Duncan Snidal, 'Hard and Soft Law in International Governance' (2000) 54(3) International Organization 421; Bryan Druzin, 'Why Does Soft Law Have any Power Anyway?' (2017) 7(2) Asian Journal of International Law 361.

There is a welcome degree of international convergence around key principles for the deployment and use of AI—particularly transparency, accountability, data privacy and protection, fairness, inclusivity, and the prevention of bias and discrimination.[34] Not surprisingly, there is no convergence around specific definitions and implementation criteria for these principles. Indeed, national approaches to these issues depend on the respective economic, business, and technological landscape. Some countries adopt a 'laissez-faire' approach, avoiding any interference with technological innovation, while others may opt for stringent regulations and broader definitions of the AI systems that fall under them. The challenge will be to promote cooperation among jurisdictions in the form of bilateral or multilateral agreements on data sharing, the development of common principles and standards on AI, and coordinated action around the enforcement of rules when AI systems cross international boundaries.[35] Such cooperation should be embedded throughout the domestic rulemaking context, which should consider existing international instruments, assess impacts beyond borders, and streamline mutual recognition of procedures.[36]

Regarding oversight of AI, governments may entrust it to dedicated agencies or departments, either new or evolved. These agencies may be responsible for developing policies and guidelines, monitoring compliance, and enforcing regulations. Some jurisdictions may opt for the model of industry self-regulation, where industry associations and tech companies develop their own guidelines and best practices. This form of privatisation of regulation can be beneficial under certain circumstances[37] but has been the subject of criticism for producing weak and ineffective norms due to barriers to inter-firm collective action, lenience in the face of noncompliant behaviour and free-rider problems, lack of public participation or oversight, and so on.[38]

---

[32] Gregory Shaffer and Mark Pollack, 'Hard Versus Soft Law in International Security' (2011) 52 Boston College Law Review 1147.
[33] Emer O'Hagan, 'Too Soft to Handle? A Reflection on Soft Law in Europe and Accession States' (2004) 26(4) Journal of European Integration 379; Jan Klabbers, 'The Undesirability of Soft Law' (1998) 67 Nordic Journal of International Law 381.
[34] Jessica Fjeld and others, 'Principled Artificial Intelligence: Mapping Consensus in Ethical and Rights-based Approaches to Principles for AI' (Berkman Klein Center for Internet & Society Report, 2020).
[35] Peter Cihon, 'Standards for AI Governance: International Standards to Enable Global Coordination in AI Research and Development' (Center for the Governance of AI, Future of Humanity Institute, Oxford, Technical Report, 2019) <www.governance.ai/research-paper/standards-for-ai-governance-international-standards-to-enable-global-coordination-in-ai-research-development> accessed 14 September 2025.
[36] OECD, 'International Regulatory Co-operation' (OECD Best Practice Principles for Regulatory Policy, 2021).
[37] Margot Priest, 'The Privatization of Regulation: Five Models of Self-Regulation' (1998) 29(2) Ottawa Law Review 233.
[38] William Bendix and Jon MacKay, 'Fox in the Henhouse: The Delegation of Regulatory and Privacy Enforcement to Big Tech' (2022) 30(2) International Journal of Law and Information

A better alternative would be a hybrid model of meta-regulation, which favours collaboration and interaction among government agencies, industry, and other stakeholders to develop policies and binding rules.[39]

To achieve these goals in a supranational context, a range of supervisory mechanisms are available.[40] The simplest option would be to follow a national supervisory model supported by the harmonisation of certain substantive rules at supranational level. In this model, national supervisory authorities would supervise the application of harmonised rules on AI within their jurisdictions, emphasising independence and cooperation among themselves, without the oversight or even coordination of a supranational body. A second option would be to introduce a two-layer framework with national supervisory authorities and a supranational body, which would play simply an advisory and coordinating role.[41] A third option would be to establish a strengthened supranational agency and entrust it with significant supervisory powers, such as direct supervision of certain entities and activities.

Regardless of the model used, the key challenge is the evolving technological landscape, which often outpaces regulatory efforts. Regulators and supervisors will continue to struggle to keep up with new AI applications, while their initiatives may have unintended consequences, such as restraining innovation or creating excessive compliance burdens and costs. The concept known as the law of unintended consequences is frequently mentioned, but consistently disregarded by politicians and prevailing public sentiment.[42] In the context of information technology, poorly-designed regulation may result in undesirable and unforeseen consequences, such as de-risking, barriers to entry, excessive compliance burdens, adverse effects on competition, and privacy and security concerns.[43] Therefore, policymakers must strike the right balance between innovation and regulation without

---

Technology 115; Ian Maitland, 'The Limits of Business Self-Regulation' (1985) 27(3) California Management Review 132.

[39] Cary Coglianese and Evan Mendelson, 'Meta-Regulation and Self-Regulation' in Robert Baldwin, Martin Cave, and Martin Lodge (eds), *The Oxford Handbook of Regulation* (OUP 2010); Ifeoma Elizabeth Nwafor, 'AI ethical Bias: A Case for AI Vigilantism (AIlantism) in Shaping the Regulation of AI' (2021) 29(3) International Journal of Law and Information Technology 225.

[40] Madeleine McNamara, 'Starting to Untangle the Web of Cooperation, Coordination, and Collaboration: A Framework for Public Managers' (2012) 35(6) International Journal of Public Administration 389.

[41] Georgi Gitchev, 'The Governance of the AI Act: Your Questions Answered' (*European AI Alliance Blog*, 4 March 2022) <https://futurium.ec.europa.eu/en/european-ai-alliance/blog/governance-ai-act-your-questions-answered> accessed 14 September 2025.

[42] Rob Norton, 'Unintended Consequences' (*EconLib* 2023) <www.econlib.org/library/Enc/UnintendedConsequences.html> accessed 14 September 2025.

[43] This was the case of the so-called 'Crypto-Wars' and the attempt of several countries to limit the public's access to strong cryptography, in order to facilitate decryption by national intelligence agencies; see Cian Murphy, 'The Crypto-Wars Myth: The Reality of State Access to Encrypted Communications' (2020) 49 Common Law World Review 245; Paul McLaughlin, 'Crypto Wars 2.0: Why Listening to Apple on Encryption Will Make America More Secure' (2016) 30 Temple International and Comparative Law Journal 353.

hindering investments in AI or putting businesses and tech companies in an unfavourable position vis-à-vis their global competitors.[44]

## 4. The EU AI Office is not enough: why the EU needs a full-fledged AI agency

At the EU level, the adoption of the EU AI Act promises to bring about greater clarity by ensuring that the rights and responsibilities of AI developers and users are interpreted consistently within the EU single market. The Act introduces a single, directly applicable set of rules for the development, placing on the market, and use of AI systems, replacing fragmented national approaches. The Act categorises AI systems based on the level of risk they pose (eg unacceptable risk, high-risk, limited risk), ensuring that obligations (such as conformity assessments, transparency duties, or prohibitions) are clear and proportionate to the risks involved.[45] The Act provides standardised procedures and requirements for AI systems, especially high-risk systems, including documentation, record-keeping, human oversight, and post-market monitoring. These elements promise to instil increased confidence among businesses, investors, and consumers while preventing the practice of seeking more favourable regulatory environments, at least within the EU single market.

The EU AI Office, established under the AI Act, will play a role in coordinating the implementation of the regulation, developing guidance, facilitating expert cooperation, and monitoring emerging AI trends. However, it is not an independent regulatory body but operates within the European Commission, with limited powers in direct supervision, enforcement, and binding decision-making. It lacks the legal autonomy and institutional strength of fully fledged EU agencies such as the European Data Protection Supervisor (EDPS), the European Medicines Agency (EMA), or the European Securities and Markets Authority (ESMA). As such, while it represents an important step forward, it does not close the debate on whether a dedicated, supranational EU AI agency—with greater independence, enforcement authority, and cross-border supervisory capacity—may still be necessary to ensure uniform application of the AI Act, safeguard fundamental rights,[46] and strengthen Europe's strategic position in global AI governance.

While the AI Act champions foundational EU values like democracy, fundamental rights, and the rule of law, its success hinges on translating these principles into practice.[47] The AI Office's integration within the European Commission risks its functional independence and raises concerns about politicisation. Furthermore, its bureaucratic structure may stifle the dynamic stakeholder engagement essential for agile governance, as participation is constrained by standard Commission procedures rather than flexible, purpose-built channels. These institutional shortcomings highlight the advantages of a structurally independent agency. Legislators naturally seek to control agencies, but overly tight reins—meant to ensure accountability—can ultimately cripple an agency's capacity to develop effective policy solutions.[48] Several factors underscore the need for a more powerful supranational EU agency dedicated to overseeing certain AI-related matters. The first factor is the inherently transnational nature of AI technologies. AI systems, by design, operate beyond the confines of national borders, coupled with the fact that data flow across countries almost without restrictions. In this context, the impact of AI applications is global and not limited to specific jurisdictions. Second, the stakes are high: AI poses serious ethical and societal challenges, security risks, and human rights risks, thereby requiring a collective approach at the level of standard-setting and enforcement. Indeed, a patchwork of national regulatory mechanisms would lead to fragmentation and harmful regulatory competition (a race to the bottom), allow forum shopping, and undermine the effectiveness of regulation.[49] Regarding oversight and enforcement, national interventions on AI uses by public authorities would be needed, but the inevitable differences in the responsibilities and traits of national supervisors might produce inconsistent levels of quality and efficacy in AI supervision across the EU, even if substantive rules are effectively harmonized. These challenges necessitate a more integrated and supranational supervisory structure, based on a centralization model that guarantees uniform enforcement, enables better coordination, and curbs forum shopping.[50]

---

[44] Georgios Pavlidis, 'Europe in the Digital Age: Regulating Digital Finance without Suffocating Innovation' (2021) 13(2) Law, Innovation and Technology 464.
[45] Nicoletta Rangone and Luca Megale, 'Risks Without Rights? The EU AI Act's Approach to AI in Law and Rule-Making' (2025) European Journal of Risk Regulation 1.
[46] Francesca Palmiotto, 'The AI Act Roller Coaster: The Evolution of Fundamental Rights Protection in the Legislative Process and the Future of the Regulation' (2025) 16 European Journal of Risk Regulation 770.
[47] Nathalie Smuha and Karen Yeung, 'The European Union's AI Act: Beyond Motherhood and Apple Pie?' in Nathalie A Smuha (ed), *The Cambridge Handbook of the Law, Ethics and Policy of Artificial Intelligence* (CUP 2025).

This rationale is further supported by broader institutional developments within the EU. In the last decade, there has been a manifest trend in favour of such EU-centralised supervision and 'agencification' in the implementation of EU law.[51] In this model, EU agencies must not be seen as autonomous regulators at the federal level, such as in the case of the division of authority between federal and state bureaucracies in the United States; the EU model favours administrative networks and the development of networked institutional relations between the EU and national bodies.[52] A successful

---

[48] Berthold Rittberger and others, 'The Competence-Control Dilemma and the Institutional Design of European Union Agencies' (2024) 37 Governance 1413.

[49] Frank Biermann and others, 'The Fragmentation of Global Governance Architectures: A Framework for Analysis' (2009) 9(4) Global Environmental Politics 14.

[50] Filipe Brito Bastos and Przemysław Pałka, 'Is Centralised General Data Protection Regulation Enforcement a Constitutional Necessity?' (2023) 19 European Constitutional Law Review 487.

[51] Edoardo Chiti, 'Decentralized Implementation: European Agencies' in Robert Schütze and Takis Tridimas (eds), *Oxford Principles of European Union Law: The European Union Legal Order* (OUP 2018) 748–776; Mira Scholten, Marloes van Rijsbergen, 'The Limits of Agencification in the European Union' (2015) 15(7) German Law Journal 1223; Takis Tridimas, 'Financial Supervision and Agency Power: Reflections on ESMA' in Niamh Nic Shuibhne and Laurence Gormley (eds), *From Single Market to Economic Union: Essays in Memory of John A. Usher* (OUP 2012) 55.

[52] Johanes Saurer, 'Supranational Governance and Networked Accountability Structures: Member State Oversight of EU Agencies' (2011) 2(1) European Journal of Risk Regulation 51; see also Herwig Hofmann, 'Mapping the European Administrative Space' (2008) 31 West European Politics 662; Ellen Vos, 'Independence, Accountability and Transparency of European Regulatory Agencies' in Damien Geradin, Rodolphe Muñoz & Nicolas Petit (eds),

example would be the model for the prudential supervision of credit institutions by the European Central Bank, which acquired supranational powers in 2014 and now follows a direct supervision model for systemically important banks.[53]

Despite legitimate concerns regarding the expanding role of EU agencies in law enforcement, this power can be counterbalanced by safeguards, including the non-delegation doctrine,[54] procedural protections, the right to good administration, and judicial review of EU agencies' acts. However, jurisprudence such as the Meroni and ESMA short-selling cases[55] reveals that while the Lisbon Treaty advanced the legitimisation of agencies, the precise scope of their powers remains critically undefined. The European Court of Justice's lenient approach in ESMA short selling failed to provide clarity or to compel the Union Legislator and Member States to define jurisdictional boundaries.[56] This persistent legal ambiguity will likely fuel further agencification, making it imperative to mitigate the accompanying risks of a deepening democratic deficit and unresolved accountability gaps.

The establishment of the EU AI Office is a step in the right direction. However, we argue that its mission should not be limited to supporting and coordinating national supervisors; it should also include policy development, standard-setting, monitoring, enforcement, and direct supervision for systematically important entities. Closely related is the task of threat assessment and risk management, which can provide valuable insights into the evolving risks of AI applications and thus help shape regulatory decisions.[57] Indeed, even after the adoption of the EU AI Act, harmonising AI regulations will remain a complex and unremitting task. To keep pace with new opportunities and risks, the new EU agency on AI would have to opt for

---

*Regulation through Agencies in the EU* (Edward Elgar 2005) 120; Wolfgang Weiß, 'Agencies versus Networks: From Divide to Convergence in the Administrative Governance in the EU' (2009) 61 Administrative Law Review 45.
[53] Less significant institutions, which are the vast majority of euro area banks, are supervised by national competent authorities (NCAs), under ECB oversight (indirect supervision); see Gianni Lo Schiavo, 'The Single Supervisory Mechanism (SSM) and the EU Anti-Money Laundering Framework Compared: Governance, Rules, Challenges and Opportunities' (2022) 23 Journal of Banking Regulation 91. It has been correctly pointed out how existing EU agencies can serve as a model facilitating the copying of institutional trends from other policy domains; Laurens van Kreij, 'How Have EU Legislators Established EU Agencies with Enforcement Tasks? Case Studies of the European Aviation Safety Agency and the European Medicines Agency' (2025) 63 Journal of Common Market Studies 590.
[54] Marta Simoncini, *Administrative Regulation beyond the Non-Delegation Doctrine: A Study on EU Agencies* (Hart 2018).
[55] Cases C-9/56 & C-10/56 *Meroni & Co, Industrie Metallurgiche, SpA v High Authority of the European Coal and Steel Community* ECLI:EU:C:1958:7; Case C-270/12 *United Kingdom v European Parliament and Council* ECLI:EU:C:2014:18.
[56] Miroslava Scholten and Marloes van Rijsbergen, 'The Limits of Agencification in the European Union' (2014) 15 German Law Journal 1223.
[57] Martin Lundgren and Ali Padya, 'A Review of Cyber Threat (Artificial) Intelligence in Security Management' in T Sipola, T Kokkonen, and M. Karjalainen (eds), *Artificial Intelligence and Cybersecurity* (Springer 2023); Nikolaos Doukas, Peter Stavroulakis, and Nikolaos Bardis, 'Review of Artificial Intelligence Cyber Threat Assessment Techniques for Increased System Survivability' in M Stamp, M Alazab, and A Shalaginov (eds), *Malware Analysis Using Artificial Intelligence and Deep Learning* (Springer 2021).

an agile and adaptable approach to standard-setting.[58] This would include iterative and flexible assessment cycles and updates to the standards, using technological solutions to improve the quality of evidence, encouraging public and stakeholder engagement, and employing non-legally binding approaches as an alternative or complement to traditional regulatory instruments.[59] The new agency should also be involved in the authorisation of 'regulatory sandboxes', which allow for reduced regulatory requirements and regulatory waivers and enable firms to test new technology models.[60] A new and strengthened EU agency on AI could also play a role in fostering research and innovation and providing capacity building and training. In this context, the ENISA model could also be used as a point of reference.[61] Its mandate would extend beyond regulatory sandboxes to include proactive measures like real-time audits, mandatory incident reporting, and EU-wide risk mapping, producing assessments more comprehensive and timelier than any national body could achieve. Centralising data would enable cross-sectoral learning, while harmonised guidelines and direct SME support would lower compliance costs and foster a truly integrated market for AI.

The structure of a new and strengthened EU agency on AI would need to include various components, to ensure its independence and effectiveness. First, a governing board comprising representatives from Member States, in particular, representatives from national supervisory authorities, as well as AI experts, would be entrusted with providing policy direction and ensuring alignment with EU values and principles. The board would also be responsible for all biding decisions on directly supervised entities or decisions regarding national supervisory authorities. Second, establishing an independent administrative board of review, following the model of the European Central Bank and the EU Anti-Money Laundering Authority (AMLA),[62] would allow the handling of appeals against binding decisions in a manner that is complementary (ie not alternative) to court proceedings. Third, specialised technical committees could focus on specific domains, such as the use of AI in finance and criminal justice. These committees would be entrusted with the development of sector-specific standards and guidelines, which would be endorsed by the governing board, following the model of AMLA. Fourth, an administrative unit would be required to manage day-to-day operations, budget issues, and coordination with Member States and other stakeholders. In this context, an executive director would manage the administrative unit and represent the AI agency. There would also be the need for mechanisms to ensure accountability and public scrutiny of the agency's organs and activities. This could be achieved through the organisation of compliance assessments and regular reporting, such as to the European Parliament and the Council, which are often called upon to ensure transparency, integrity, and accountability at the EU level.[63] This would address growing concerns about the accountability of EU-level agencies, which are often viewed as powerful and unaccountable bureaucracy.[64]

---

[58] Wendell Wallach and Gary Marchant, 'An Agile Ethical/Legal Model for the International and National Governance of AI and Robotics' (Proceedings of the 2018 AAAI / ACM Conference on Artificial Intelligence, Ethics and Society, 2018) <www.aies-conference.com/2018/contents/papers/main/AIES_2018_paper_77.pdf> accessed 14 September 2025.
[59] OECD, 'Practical Guidance on Agile Regulatory Governance to Harness Innovation', OECD Regulatory Policy Committee, GOV/RPC(2021)14/FINAL, 2021.
[60] OECD, 'The role of sandboxes in promoting flexibility and innovation in the digital age' (Going Digital Toolkit Policy Note, 2020) <www.oecd.org/en/publications/the-role-of-sandboxes-in-promoting-flexibility-and-innovation-in-the-digital-age_cdf5ed45-en.html> accessed 14 September 2025.
[61] Dimitra Markopoulou, Vagelis Papakonstantinou, and Paul de Hert, 'The New EU Cybersecurity Framework: The NIS Directive, ENISA's Role and the General Data Protection Regulation' (2019) 35(6) Computer Law & Security Review 105336; Jukka Ruohonen, Sami Hyrynsalmi, and Ville Leppänen, 'An Outlook on the Institutional Evolution of the European Union Cyber Security Apparatus' (2016) 33(4) Government Information Quarterly 746.
[62] Georgios Pavlidis, 'The Birth of the New Anti-money Laundering Authority: Harnessing the Power of EU-wide Supervision' (2024) 31 Journal of Financial Crime 322.

The establishment of a strengthened EU agency on AI should not lead to duplication, overlap, or inconsistencies with existing regulatory bodies, which would do more harm than good by accentuating legal uncertainties.[65] Therefore, the new AI agency would coordinate closely with existing bodies, such as ENISA, the European Data Protection Board and the European Union Agency for Fundamental Rights, and sector-specific bodies and would streamline rules and procedures through collaborative efforts. A more complex challenge involves structuring coexistence and coordination between a new EU AI agency and national regulatory bodies to prevent conflicts and ensure a uniform regulatory playing field. This is critical because the implementation of the AI Act is a shared responsibility, involving both established EU institutions like the Commission and new bodies like the AI Office, alongside national authorities.[66] The coordination problem is especially acute in Member States where national supervisory authorities lack full independence. Here, a supranational agency could act as a crucial counterweight, guaranteeing consistent enforcement while accommodating diverse national traditions. To bridge these institutional differences, formalised cooperation mechanisms, such as joint supervisory teams, mandatory peer reviews, and binding mediation, would be indispensable.

The EU agency on AI should have the powers to ensure that national bodies abide by the same high standards because high-quality regulation at a certain level of governance can be compromised by poor regulatory policies and practices at other levels in multi-level regulatory governance models.[67] The model of AMLA could be followed in this context, in which national authorities supervise certain activities and entities, while the supranational agency focuses on supervising high-risk entities; it also supports national authorities and promotes supervisory convergence. Therefore, direct EU supervision is applied only when there is evidence that national action alone is insufficient. This is consistent with the principle of subsidiarity, which recognises that national supervisory authorities will not be removed but will become part of an integrated supervisory system, even if direct supervision for some entities is transferred to the EU.[68] This is also consistent with the principle of proportionality, which entails giving adequate but not excessive authority and resources to EU bodies.[69] Finally, a strengthened EU agency on AI would also support mutually beneficial public–private partnerships[70] as well as an open and continuous dialogue with the AI industry, civil society organisations, and academia, following the model of the Financial Action Task Force and its dialogue with the private sector and other stakeholders in the field of anti-money laundering.[71]

---

[63] European Parliament, 'Transparency, Integrity and Accountability in the EU Institutions', Briefing, European Parliament's Committee on Petitions, 2019.
[64] Madalina Busuioc, *European Agencies: Law and Practices of Accountability* (OUP 2013).
[65] Vera Lúcia Raposo, 'Ex Machina: Preliminary Critical Assessment of the European Draft Act on Artificial Intelligence' (2022) 30(1) International Journal of Law and Information Technology 88.
[66] Claudio Novelli and others, 'A Robust Governance for the AI Act: AI Office, AI Board, Scientific Panel, and National Authorities' (2025) 16 European Journal of Risk Regulation 566.
[67] Delia Rodrigo, Lorenzo Allio, and Pedro Andres-Amo, 'Multi-Level Regulatory Governance: Policies, Institutions and Tools for Regulatory Quality and Policy Coherence', OECD Report, 2009.

Most importantly, a strengthened EU agency on AI, in contrast to the EU AI Office, would need to have functional and budgetary independence which is a key issue for new agencies. The EU would have to ensure that adequate funding, staff, and expertise are channelled to the new body to support its operations. Funding sources may include a combination of fees paid by industry stakeholders, contributions from Member States, and funding from the EU budget. Furthermore, the new agency would have functional autonomy – which means it would enjoy decision-making discretion and be able to resolve policy or managerial issues without external interference – although interplay and consultations between actors at various levels, including the European Commission, would still be conceivable.[72] In the context of regulations establishing new European supervisors, the independence principle is consistently adopted to ensure that independence problems at the national level are not transferred to the EU level or vice versa.[73] Similar means of oversight, based on independence and expertise, align perfectly with complex and rapidly advancing domains such as AI.[74]

Nevertheless, there are concerns and objections to be addressed. A recurring theme in the process of European integration has been the fear of the potential loss of sovereignty.[75] The field of AI will be no exception. Indeed, Member States may consider that relinquishing considerable control over AI regulation and oversight to a supranational EU agency will limit their ability to address AI-related issues in accordance with national priorities and preferences. As with many EU initiatives, the optimal solution is a body designed for collaborative governance with Member States, integrating national perspectives into AI policy while respecting subsidiarity and proportionality. Critically, any such agency must align with the Court of Justice's jurisprudence on delegated powers. While the aforementioned Meroni and ESMA short-selling doctrines prohibit delegating wide discretionary powers without adequate safeguards, they permit delegation when powers are clearly circumscribed, subject to judicial review, and necessary to achieve Treaty objectives. A carefully designed AI Agency, with structured accountability and oversight, would operate firmly within these established legal boundaries.

---

[68] Christoph Henkel, 'The Allocation of Powers in the European Union: a Closer Look at the Principle of Subsidiary' (2002) 20 Berkeley Journal of International Law 359; Gabriel Moens, John Trone, 'The Principle of Subsidiarity in EU Judicial and Legislative Practice: Panacea or Placebo' (2015) 41 Notre Dame Law School Journal of Legislation 65; Oxana Pimenova, 'Subsidiarity as a regulation principle in the EU' (2016) 4(3) The Theory and Practice of Legislation 381.

[69] Merijn Chamon, *EU Agencies: Legal and Political Limits to the Transformation of the EU Administration* (OUP 2016); Darren Harvey, 'Federal Proportionality Review in EU Law: Whose Rights are they Anyway?' (2020) 89(3-4) Nordic Journal of International Law 303; see also Wolf Sauter, 'Proportionality in EU Law: A Balancing Act?' (2013) 15 Cambridge Yearbook of European Legal Studies 439.

[70] Nutavoot Pongsiri, 'Regulation and Public-private Partnerships' (2002) 15(6) International Journal of Public Sector Management 487.

[71] Mark Seidenfeld, 'Empowering Stakeholders: Limits on Collaboration as the Basis for Flexible Regulation' (1999) 41 William and Mary Law Review 411.

[72] Per Lægreid, Koen Verhoest, and Werner Jann, 'The Governance, Autonomy and Coordination of Public Sector Organizations' (2008) 8 Public Organization Review 93.

[73] Annetje Ottow, 'Independent Supervisory Authorities: A Fragile Concept' (2012) 39(4) Legal Issues of Economic Integration 419.

[74] Michelle Everson, 'Independent Agencies: Hierarchy Beaters?' (1995) 1(2) European Law Journal 180.

Furthermore, concerns regarding bureaucratic inefficiency and red tape constitute a popular perception and recurrent point of criticism against the EU.[76] For this reason, the design of a new and strengthened EU agency on AI should prioritise the principles of agility and responsiveness, eg ensure a meaningful and timely role in the development of draft technical standards. This institutional design choice corresponds to the mission of such an EU agency, that is, to solve the cooperation problems that international actors face in the specific area of AI governance.[77]

The institutional foundation of such an agency could rest on Articles 114 and 352 TFEU, with the latter serving as a flexibility clause and fallback legal basis. This approach has a clear precedent: Article 114 was the basis for the European Banking Authority, while Article 352 (formerly Article 308 EC) was used for the European Union Agency for Fundamental Rights. Establishment would require a new Regulation, adopted under the ordinary legislative procedure, to amend the AI Act, thereby guaranteeing the full democratic involvement of both the European Parliament and the Council. The successful precedent of the AMLA proves that even with Treaty constraints on the delegation of powers, a supranational agency can be granted robust authority if it is constructed with sufficient accountability mechanisms and review procedures.

---

[75] Raffaele Bifulco and Alessandro Nato, 'The Concept of Sovereignty in the EU: Past, Present and the Future' (RECONNECT – Reconciling Europe with its Citizens through Democracy and Rule of Law, Working Paper, 2020; Ole Waever, 'Identity, Integration and Security: Solving the Sovereignty Puzzle in EU Studies' (1995) 48(2) Journal of International Affairs 389; Neil MacCormick, 'The Maastricht-Urteil: Sovereignty Now' (1995) 1(3) European Law Journal 259; Martin Loughlin, 'Why Sovereignty?' in Richard Rawlings, Peter Leyland, and Alison Young (eds), *Sovereignty and the Law: Domestic, European and International Perspectives* (OUP 2013) 34–49.
[76] Wim Voermans and others, 'Codification and Consolidation in the European Union: A Means to Untie Red Tape' (2008) 29 Statute Law Review 65.
[77] This corresponds to the model proposed by Barbara Koremenos, Charles Lipson and Duncan Snidal, 'The Rational Design of International Institutions' (2001) 55(4) International Organization 761.

## 5 Concluding remarks

As AI technologies advance, numerous jurisdictions are likely to establish their own AI agencies, which will inevitably add to the complexity of global AI governance.[78] The EU must engage with international partners and help create global AI standards and shared ethical frameworks and principles, hopefully with the positive contribution of the United Nations and the G20.[79] This will ensure the interoperability of AI systems and foster innovation and international trade in AI products and services. Nevertheless, as mentioned earlier, there is no universally agreed-upon set of standards for AI, which leaves the door open for further regulatory divergence.

The EU, assisted by the EU AI Office, or a future strengthened EU agency on AI, should promote international collaboration in the form of bilateral or multilateral agreements that deal with information exchanges, mutual administrative and judicial assistance in cross-border cases, data sharing, common responses to AI-related threats, etc.[80] A strengthened EU agency on AI should be seen as a part of the future international AI ecosystem, which would gradually be enriched with specialised regulatory and supervisory bodies in many countries. In this environment, this agency would be the voice of the EU in the global collaborative effort to develop AI standards and ethical frameworks and collectively address AI and cybersecurity threats.[81]

Nevertheless, it is incumbent upon the EU to exercise circumspection in this context. While a collaborative vision of global AI governance is an ethically sound and justified approach, sober assessment dictates that one must expect significant global antagonism and a strenuous race to leverage AI, as numerous countries are poised to endorse strategic policies and commit significant financial investment to catalyse innovation in this field. In this context, the EU must be vigilant against serious risks, such as instances of intellectual property theft and science espionage, cross-border data migration towards jurisdictions with less stringent data protection regulations, global startup ecosystem competition with other countries, and competition relating to incentives and public expenditure on research and development.[82] Therefore, the EU must develop effective mechanisms to elevate its 'technological sovereignty'.[83] In practice, this means that the EU must avoid relying on a limited number of third-country suppliers to obtain technologies, such as AI, which are essential for startups and the EU economy in general.[84] If designed correctly, a strengthened EU agency on AI could deliver on this promise by accelerating a unified approach to AI regulation and promoting innovation through knowledge dissemination and resource pooling. For this reason, establishing such an agency should be a matter of priority.

---

[78] Karen Alter, Kal Raustiala, 'The Rise of International Regime Complexity' (2018) 14(1) Annual Review of Law and Social Science 329.

[79] Eugenio Garcia, 'Multilateralism and Artificial Intelligence: What Role for the United Nations?' in Maurizio Tinnirello (ed), *The Global Politics of Artificial Intelligence* (Routledge 2020) 1-20; Thorsten Jelinek, Wendell Wallach & Danil Kerimi, 'Policy Brief: The Creation of a G20 Coordinating Committee for the Governance of Artificial Intelligence' (2021) 1 AI and Ethics 141.

[80] Eleonore Pauwels, 'The New Geopolitics of Converging Risks: The UN and Prevention in the Era of AI' (United Nations University, Centre for Policy Research, Technical Report, 2019) <https://i.unu.edu/media/cpr.unu.edu/attachment/3472/PauwelsAIGeopolitics.pdf> accessed 14 September 2025.

[81] Alan Bundy, 'Preparing for the Future of Artificial Intelligence' (2017) 32 AI & Society 285.

[82] Wolfgang Dierker, 'Technologische Souveränität: Begriff und Voraussetzungen im transatlantischen Kontext' (2023) 103(6) Wirtschaftsdienst 386; see also European Parliament, 'Key Enabling Technologies for Europe's Technological Sovereignty' (European Parliamentary Research Service Study, 2021).

Despite rapid advancements in new technologies, the development of legal frameworks often lags behind, mainly due to the complexity of the legislative process and the need for consensus among multiple actors.[85] We argue that there is a need to expedite EU efforts to establish supranational oversight. A strengthened EU AI agency could provide significant benefits by averting the emergence of a fragmented and decentralised landscape characterised by varying approaches and enforcement attitudes among EU Member States and, more broadly, different geographical and cultural clusters.[86] With the EU AI Act, there has been a shift from soft-law principles to hard-law regulations, which is timely and appropriate in the context of AI. The next logical step is the establishment of a robust supranational authority on AI with powers and responsibilities in threat assessment, standard-setting, supervision, and enforcement. Of course, concerns about loss of sovereignty and bureaucratic inefficiency should be addressed. This could be achieved through collaboration with Member States, agile agency design, transparency, and public accountability mechanisms.[87] Finally, a new and strengthened agency should have access to the necessary resources (funding, staff, technology) and, most importantly, enjoy a high degree of independence to accomplish its mission and promote close coordination with public and private stakeholders and public bodies, both at the EU and national levels, to avoid duplication or inconsistency of efforts.

---

[83] Although there are various interpretations, the term generally refers to a form of collective control of digital content and/or infrastructures; Stephane Couture, Sophie Toupin, 'What Does the Notion of "Sovereignty" Mean when Referring to the Digital?' (2019) 21(10) New Media & Society 2305; Francesco Crespi and others, 'European Technological Sovereignty: An Emerging Framework for Policy Strategy' (2021) 56(6) Intereconomics 348.

[84] European Innovation Council, 'Statement on Technological Sovereignty' (Annex to the Statement of the EIC pilot Advisory Board at the launch of the EIC, 2021) <https://eic.ec.europa.eu/system/files/2021-03/EIC%20Advisory%20Board%20statement%20at%20launch%20of%20EIC_1.pdf> accessed 14 September 2025.

[85] Gary Marchant, Branden Allenby, and Joseph Herkert (eds), *The Growing Gap Between Emerging Technologies and Legal-ethical Oversight: The Pacing Problem* (Springer 2011).

[86] Lewin Schmitt, 'Mapping Global AI Governance: A Nascent Regime in a Fragmented Landscape' (2022) 2 AI Ethics 303; Niels van Berkel and others, 'A Systematic Assessment of National Artificial Intelligence Policies: Perspectives from the Nordics and Beyond' in NordiCHI '20 Proceedings of the 11th Nordic Conference on Human-Computer Interaction: Shaping Experiences, Shaping Society' 2020.

[87] Marijn Janssen and others, 'Transparency-by-design as a Foundation for Open Government' (2017) 11(1) Transforming Government: People, Process and Policy 2; Madalina Busuioc, 'European Agencies and their Boards: Promises and Pitfalls of Accountability Beyond Design' (2012) 19(5) Journal of European Public Policy 719.